# Unitary Solution to a Quantum Gravity Information Paradox

T. C. Ralph

Department of Physics, University of Queensland, Brisbane 4072, QLD, Australia
(Dated: October 24, 2018)

We consider a toy model of the interaction of a qubit with an exotic space-time containing a time-like curve. Consistency seems to require that the global evolution of the qubit be non-unitary. Given that quantum mechanics is globally unitary, this then is an example of a quantum gravity information paradox. However, we show that a careful analysis of the problem in the Heisenberg picture reveals an underlying unitarity, thus resolving the paradox.

## I. INTRODUCTION

An intriguing point of conflict between general relativity and quantum mechanics is the question of unitarity. For example, if Hawking radiation [1] is truly thermal, as predicted by semi-classical general relativity, then the complete evaporation of a black-hole would imply the loss of all the information that has passed into it. This is clearly a non-unitary evolution. In contrast quantum mechanics is globally unitary. Several possible solutions to this problem have been put forward focusing on the black hole example. These can be grouped into two camps: those that argue that a full theory of quantum gravity will enable information to escape from black holes, thus restoring unitarity [2] and; those that argue that such a full theory will inevitably be non-unitary [3]. It has recently been shown that there is no "easy fix" to the problem in terms of hidden correlations [4]. In this paper we use a toy model to suggest a third alternative: that from the usual point of view the information *is* lost, however if we allow exotic measurement techniques it can be recovered, demonstrating that the fundamental evolution is unitary. The key physics in our approach is that we consider different space-time histories to be distinguishable and allow them to add coherently.

Perhaps the simplest example of a model combining general relativistic and quantum mechanical evolutions that naturally leads to non-unitarity was introduced by Deutsch [5] and later developed by Bacon [6]. The Deutsch-Bacon (D-B) model assumes that time-like curves can be created, as predicted by the metric introduced by Morris et al [7]. Time-like curves allow a particle to follow a trajectory into its own past. The model considers a broad class of interactions between two-level quantum systems (qubits) and time-like curves and shows that, whilst self consistent solutions are always possible, the solutions are in general non-unitary. An interesting feature of this model is that, if we assume time-like curves can exist [8], then we can solve the resulting evolutions consistently without requiring a full theory of quantum gravity. In contrast, the final stages of evaporation of a black hole would appear to intrinsically require a full theory of quantum gravity.

## II. THE D-B MODEL

Consider the situation depicted in Fig.1(a). Two qubits are involved. Qubit 1 is free propagating. Qubit 2 is trapped on a closed time-like curve formed by two mouths of a wormhole. The wormhole acts as a time machine [7]. Qubit 2 enters the future mouth of the wormhole at time $t$ and emerges at an earlier time $t' = t - \tau$ from the past mouth of the wormhole. It then propagates forward in time for $\tau$ till $t$ where it completes the causal loop by entering the future mouth of the wormhole. Qubit 1 is allowed to interact with qubit 2 via an arbitrary unitary interaction $U$. No other interactions are allowed with qubit 2. This is indicated on the figure by the dotted box. We desire the final state of qubit 1, after the interaction.

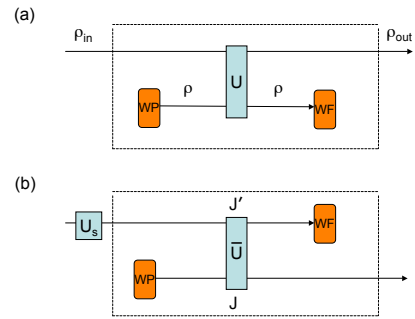

FIG. 1: Two representations of the scattering of a qubit from a wormhole with time running horizontally. In the original representation, (a), the incoming qubit is scattered from a second qubit that is trapped on a closed time-like curve by the wormhole. In the second representation, (b), the incoming qubit is sent into the past by the wormhole where it "interacts with itself". The representations are equivalent if we take $\bar{U} = U U_{swap}$. WF is the future mouth of the wormhole and WP is the past mouth of the wormhole.

D-B propose the following formalism to treat this situation and show that it always leads to a consistent solution: (i) determine the state of qubit 2, given by the

density operator $\rho$, via the consistency equation

$$\rho = Tr_1[U(\rho_{in} \otimes \rho)U^\dagger] \qquad (1)$$

where $\rho_{in}$ is the initial state of qubit 1 and the trace is over the Hilbert space of qubit 1. Given a solution for $\rho$, the output state of qubit 1, $\rho_{out}$, is given by

$$\rho_{out} = Tr_2[U(\rho_{in} \otimes \rho)U^\dagger] \qquad (2)$$

where the trace is now over the Hilbert space of qubit 2. In general the solution for $\rho_{out}$ will be non-unitary and a nonlinear function of the input state. For example, if we take the unitary to be a Controlled-not gate followed by a Swap gate, i.e. $U = U_{cnot}U_{swap}$, and consider a pure input state, $\alpha|0\rangle + \beta|1\rangle$, we find, using Eq.1 and 2, that the density operator of the output state is given by

$$\rho_{out} = (|\alpha|^4| + |\beta|^4|)|0\rangle\langle 0| + 2|\alpha\beta|^2|1\rangle\langle 1|. \qquad (3)$$

The solution is clearly non-unitary as $\rho_{out}$ is diagonal and hence all coherences have been lost. The output is also a non-linear function of the input state, a point we will return to at the end of this paper.

To recap, we see that the D-B model predicts a pure input state can be scattered into a mixed output state. Even in principle, we cannot access qubit 2 after we pass the time $t$ at which it enters the future mouth of the wormhole, and hence it would seem information about the scattered qubit is truly lost. The non-unitarity is explicitly put into the model via the traces in Eqs 1 and 2. The traces are required in order to guarantee that consistent solutions can always be found, however it would be more significant if a physical mechanism for the non-unitary behaviour could be identified. Here we introduce a different analysis of the D-B model that reveals such a mechanism. In the appropriate limits we reproduce several examples of the non-unitary evolutions of D-B. The calculations are performed in the Heisenberg Picture and do not involve any explicit traces. In fact, it is found that when interactions are chained in the appropriate way the evolution *is* in fact reversible. Thus the non-unitarity is only apparent and the evolutions can be understood as fundamentally unitary processes. We speculate that this effect may be more general than this specific model.

### III. THE HEISENBERG MODEL

An equivalent formulation of the problem of Fig.1(a) is shown in Fig.1(b). Now an additional swap gate is implicitly included in the arbitrary unitary $\bar{U}$, i.e. the two circuits (Fig.1(a) and (b)) are exactly equivalent if $\bar{U} = UU_{swap}$. The time-like curve is no longer explicitly closed and in some sence there is now only one qubit. The region of time in which two qubits are present can be understood as the past and present representations of a single qubit. Again, the dotted box demarks the region in which interactions with the qubit are forbidden. In the Appendix we contrive a space-time representation of the circuit of Fig.1(b) that explicitly prevents acausal effects for an external observer.

We can solve this alternative formulation in the Heisenberg Picture by considering the evolution of the Pauli operators $X, Y$ and $Z$, representing a complete set of observables for the qubit. We take the initial state to be $|0\rangle$ and explicitly include state preparation in the evolution. Expectation values can be evaluated by evolving the observables back in time through the circuit from the measurement point to the initial state [9]. The evolution is evaluated in the free-fall reference frame of the qubit. In order to do this we need to introduce an explicit temporal parameterization of the observables. We take

$$J(t_0) = \int dt \, G(t - t_0) \, J_t \qquad (4)$$

where $J = X, Y, Z$, $J_t$ is an idealized, instantaneous Pauli operator acting at time $t$ and $G(t - t_0)$ is a distribution function describing the uncertainty in when the operator acts. We assume that the standard deviation, $d$ of the distribution function $G$ is much smaller than $\tau$, such that on the scale of Fig.1(b) the qubit behaves as a point particle. We also assume that the same distribution function describes the final measurement, the collision unitary and the initial state preparation. In the reference frame of the qubit two collisions occur, one before entering the wormhole and the other after leaving it. These collisions are both *with itself at different times*. The parameter $t_0$ acts as a clock in the reference frame of the qubit and allows us to make mathematical sence of this strange statement. We now apply this method to two example $\bar{U}$'s, that illustrate the method and demonstrate its ability to reproduce the results of the original model.

In the first example we consider $U = U_{cs}U_{swap}$ and hence $\bar{U} = U_{cs}$ where $U_{cs}$ is the unitary transformation corresponding to a controlled sign. For simplicity we take the input state in the Schrödinger Picture to be the pure state $\alpha e^{i\theta}|0\rangle + \beta e^{-i\theta}|1\rangle$ with $\alpha$ and $\beta$ real. In the Heisenberg Picture we explicitly create this state using the unitary transformation $U_s = e^{i\theta Z}(\alpha I + i\beta Y)$. It is straightforward to generalize the treatment to include mixed states. We begin with the D-B model treatment. The consistency requirement (Eq.1) leads to the solution [6]

$$\begin{aligned}\rho &= \alpha^2|0\rangle\langle 0| + \beta^2|1\rangle\langle 1| \\ &+ (\alpha^2 - \beta^2)\alpha\beta(e^{-i2\theta}|0\rangle\langle 1| + e^{i2\theta}|1\rangle\langle 0|)\end{aligned} \qquad (5)$$

and hence from Eq.2 the output

$$\begin{aligned}\rho_{out} &= \alpha^2|0\rangle\langle 0| + \beta^2|1\rangle\langle 1| \\ &+ (\alpha^2 - \beta^2)^2\alpha\beta(e^{-i2\theta}|0\rangle\langle 1| + e^{i2\theta}|1\rangle\langle 0|).\end{aligned} \qquad (6)$$

As before this solution is unusual in being non-unitary and a non-linear function of the input state.

We now solve for the same unitary using the Heisenberg approach. We introduce the following nomenclature: $J = J(t), J' = J(t - \tau), J'' = J(t - 2\tau)$, etc, where $t$ is an arbitrary time and, as before, $\tau$ is the time shift produced by the worm-hole. The transformation rules for the controlled sign gate are $(I)(X) \rightarrow (Z)(X)$, $(X)(I) \rightarrow (X)(Z)$, $(I)(Y) \rightarrow (Z)(Y)$, $(Y)(I) \rightarrow (Y)(Z)$, $(I)(Z) \rightarrow (I)(Z)$ and $(Z)(I) \rightarrow (Z)(I)$, where the brackets delineate operators acting on the first and second qubit. Consider the representation of Fig.1(b). We begin with $Z$ and evolve it back from the output into the wormhole region. At the first encounter with the unitary we set the input to be $(I)(Z)$, i.e. we take the upper mode to be the identity. The evolution gives $(I)(Z) \rightarrow (I)(Z)$. Evolving the lower mode further back in time we encounter the wormhole and then arrive at the unitary again, but now in the upper mode and at an earlier time (in the qubit reference frame). We now conclude that the input to the unitary should be $(Z')(Z)$. This gives evolution through the unitary of $(Z')(Z) \rightarrow (Z')(Z)$. Propagating the lower mode back through the wormhole and to the unitary again produces the same input $(Z')(Z)$. Thus we have consistently solved the evolution through the wormhole and we conclude that the value of the upper mode that should be evolved back to the state preparation unitary, $U_s$, is $Z'$.

Now consider $X$. Evolution back through the unitary the first time gives $(I)(X) \rightarrow (Z)(X)$. Propagating the lower mode through the wormhole and back through the unitary a second time leads to the evolution $(X')(X) \rightarrow (Z)(XZ')$. A third iteration gives a consistent evolution through the unitary of $(X'Z'')(X) \rightarrow (ZX'Z'')(XZ')$ that has "converged" in the sence that further iterations do not change the evolution. We conclude that $ZX'Z''$ should be evolved back to the state preparation unitary. Similarly we find that $Y$ evolves to $ZY'Z''$. Finally we evolve the operators back through the state preparation unitary, $U_s$, to obtain the operators, $J_m$, that act on the initial zero state

$$\begin{aligned} Z_m &= Z'_i \\ X_m &= Z_i X'_i Z''_i \\ Y_m &= Z_i Y'_i Z''_i. \end{aligned} \quad (7)$$

where

$$\begin{aligned} Z_i &= (\alpha^2 - \beta^2)Z + 2\alpha\beta X \\ X_i &= \cos 2\theta((\alpha^2 - \beta^2)X + 2\alpha\beta Z) + \sin 2\theta\, Y \\ Y_i &= \cos 2\theta Y - \sin 2\theta((\alpha^2 - \beta^2)X + 2\alpha\beta Z). \end{aligned} \quad (8)$$

These are clearly unusual expressions, containing operators representing three different temporal histories. Never-the-less we can evaluate expectation values using Eq.4. Consider the simple example of a component product of $Z$ operators with different temporal histories

$$\langle ZZ' \rangle = (\alpha^2 - \beta^2)^2 + 4\alpha^2\beta^2 \int dt\, G(t-t_0)G(t-t_0-\tau) \quad (9)$$

If $\tau = 0$ then normalization of the temporal function means we obtain $\langle ZZ \rangle = 1$. However, for the condition $\tau >> d$, then $G(t-t_0)$ and $G(t-t_0-\tau)$ are effectively orthogonal (have no overlap) and $\langle ZZ' \rangle = (\alpha^2 - \beta^2)^2 = \langle Z \rangle^2$. This technique can be generalized to obtain expectation values for the operators of Eq.7.

The solutions using the D-B model and the Heisenberg model are compared via the expectation value relations $\langle J \rangle = \langle 0|J_m|0 \rangle = Tr[J\rho_{out}]$. Both approaches lead to the same solutions, namely:

$$\begin{aligned} \langle X \rangle &= (\alpha^2 - \beta^2)^2\, 2\alpha\beta \cos 2\theta \\ \langle Y \rangle &= -(\alpha^2 - \beta^2)^2\, 2\alpha\beta \sin 2\theta \\ \langle Z \rangle &= (\alpha^2 - \beta^2) \end{aligned} \quad (10)$$

As a second example we consider the case presented earlier in which $U = U_{cnot}U_{swap}$. Hence $\bar{U} = U_{cnot}$. The transformation rules for the controlled not gate are $(I)(X) \rightarrow (I)(X)$, $(X)(I) \rightarrow (X)(X)$, $(I)(Y) \rightarrow (Z)(Y)$, $(Y)(I) \rightarrow (Y)(X)$, $(I)(Z) \rightarrow (Z)(Z)$ and, $(Z)(I) \rightarrow (Z)(I)$, where the first qubit is the control. Following the procedure described previously we find for $Z$, after two iterations, the evolution through the unitary of $(Z')(Z) \rightarrow (ZZ')(Z)$ and we conclude that $ZZ'$ should be evolved back to the state preparation unitary. Something different happens when we consider $X$. We do not find a closed solution, but rather an infinite product of operators. We conclude that the evolution through the unitary is $(X'X''X'''...)(X) \rightarrow (X'X''X'''...)(XX'X''...)$ and hence conclude that $X'X''X'''...$ should be propagated back to the unitary. For $Y$ the evolution is $(Y'X''X'''...)(Y) \rightarrow (ZY'X''X'''...)(YX'X''X'''...)$ and hence $ZY'X''X'''...$ should be propagated back to the unitary. Thus we find

$$\begin{aligned} Z_m &= Z_i Z'_i \\ X_m &= X'_i X''_i X'''_i ... \\ Y_m &= Z_i Y'_i X''_i X'''_i ... \end{aligned} \quad (11)$$

The presence of infinite direct products appears pathological, never-the-less if we ignore the singular point $\alpha = \beta$ (which is also a singular point in the D-B model [6]) the expectation values converge and we obtain values for the expectation values consistent with Eq.3, namely

$$\begin{aligned} \langle X \rangle &= 0 \\ \langle Y \rangle &= 0 \\ \langle Z \rangle &= (|\alpha|^2 - |\beta|^2)^2 \end{aligned} \quad (12)$$

## IV. REVERSABILITY OF THE HEISENBERG MODEL

Although, as we have shown, the expectation values are equivalent in the two approaches, the physical pictures are quite different. In the D-B model the non-unitary evolution emerges as a mysterious requirement for preserving consistency. In the Heisenberg model the evolution involves only unitary operations. The non-unitary behaviour of the expectation values arises from the measurements. Direct measurement of the qubit observables is unable to recognize the coherences between the various histories that contribute to the results. As a consequence those coherences are implicitly traced over in the expectation values. We might expect that these coherences will play a role if we chain a pair of wormhole interactions. We now consider such a situation and show that the two models give highly divergent results.

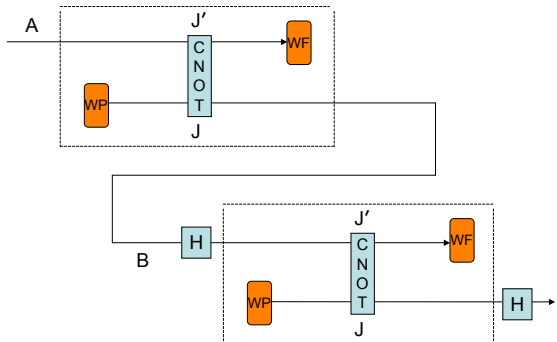

FIG. 2: Chained wormhole gates. H stands for a Hadamard gate

We consider the chained interaction depicted in Fig.2. The second wormhole interaction is the same as the first (with $\bar{U} = U_{cnot}$). We apply a Hadamard gate to the output of the first wormhole before injecting it into the second and a further Hadamard after the second wormhole. Using the D-B model we can solve this situation using Eq.1 and 2, but now with the input state $\rho_{in} = (|\alpha|^4 + |\beta|^4)|+\rangle\langle+| + 2|\alpha\beta|^2|-\rangle\langle-|$, where $|\pm\rangle = |0\rangle \pm |1\rangle$. The final output state is the completely depolarized density operator $1/2(|0\rangle\langle0| + |1\rangle\langle1|)$.

A careful analysis of the Heisenberg approach arrives at a very different conclusion. The action of the Hadamard gate is to produce the evolutions: $Z \to X, X \to Z$ and $Y \to -Y$. Combining these transformations with those of Eq. 11 we find that propagating back from the measurement point through the first wormhole to position $B$ in Fig.2 gives the transformations: $Z \to Z'Z''Z'''..., X \to XX', Y \to XY'Z''Z'''...$. Propagating back from $B$ through the unitary and wormhole gives the somewhat surprising consistent evolutions:

$$\begin{aligned}(Z''Z'''Z''''...)(Z'Z''Z'''...) &\to (Z')(Z'Z''Z'''...) \\ (X')(XX') &\to (X')(X) \\ (X'Z''Z'''...)(XY'Z''Z'''...) &\to (Y')(XZ'Z''...) \end{aligned} \tag{13}$$

where we have used the Pauli identities $JJ = I$ and $XZ = -iY$. Hence, in stark contrast to the prediction of the D-B model, we find that the Heisenberg approach predicts that the second wormhole reverses the evolution of the first wormhole, essentially giving the identity for the entire transformation, i.e. $Z \to Z', X \to X', Y \to Y'$.

## V. CONCLUSIONS

We have analyzed a simple model combining an exotic general relativistic metric with quantum mechanics that appears to inevitably lead to non-unitary evolutions. We have shown that the predictions of this model can be reproduced using a model that is explicitly unitary and in principle reversible. In the original model consistent solutions are guaranteed by ad hoc trace operations which lead to non-unitary evolution. In the new model treatment of different space-time histories through the wormhole as distinguishable paths guarantees consistent solutions and leads to apparently non-unitary evolutions. However the underlying unitarity is demonstrated by the chaining of two interactions, revealing that the evolutions are reversible.

The new model predicts expectation values that are a non-linear function of the input state as in the original model. However, the conclusions about computational complexity in the presence of these exotic gates arrived at in Ref.[6] depend on chaining of the interactions. Such chainings may not lead to the same conclusions when applied to the new model.

Although our analysis is abstract and the model rather contrived, we believe the general features of the solution may teach us about more physical situations such as the evaporation of black holes.

*Acknowledgments:* We thank G.J.Milburn, T.Downes, P.P.Rohde, A.P.Lund and C.M Savage for useful discussions.

## VI. APPENDIX

For concreteness we contrive a more explicit representation of the circuit of Fig.1(b), with a space-time geometry given by the space-time and space-space diagrams of Fig.3, that explicitly prevents acausal effects for an external observer. We assume tubes, that can only support a single quantum mode, guide our qubit into and out of the wormholes. These "wave-guides" exclude external modes from interacting with the qubit in the region between HI and HO in Fig.3. The input mouth of the wormhole is located at WI and the output at WO. The wave-guides are arranged such that an elastic collision occurs between the input and output wave-guides at point C. The collision is guaranteed to be unitary because of the exclusion of all external modes. The geometry of the wave-guide path is required by no-signaling to be such that signals cannot traverse the straight-line distance between HI and HO faster than light-speed. The coor-dinates used in Fig.3 are those of an external observer in flat space-time. We make the following (simplifying, but non-essential) assumptions about the metric experienced by a qubit in free-fall through the wave-guide. We assume the qubit sees flat space-time up till WI. In traversing WI → WO the qubit coordinates change such that $x, t \to x + \delta x, t + \delta t$ whilst the external coordinates change such that $x, t \to x + \delta x + \vec{\epsilon}, t + \delta t - \tau$ (see Fig.3). In the text we assume that $\delta x, \delta t$ are small enough to be neglected and that the free-fall speed of the qubit is sufficiently slow that special relativistic effects can be neglected.

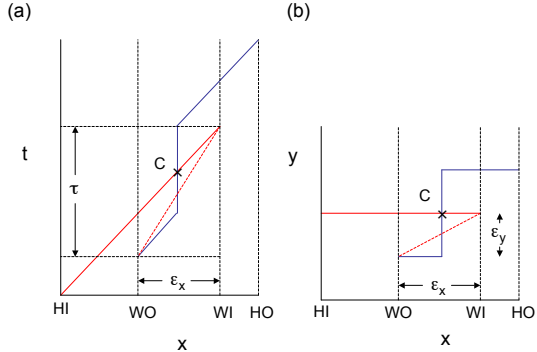

FIG. 3: Space-time (a) and space-space (b) representations of wormhole interaction. $\vec{\epsilon} = (\epsilon_x, \epsilon_y)$. See text for other definitions.